# A Low-Complexity 2D Signal Space Diversity Solution for Future Broadcasting Systems

Jianxiao YANG[1], Kai WAN[1], Benoit GELLER[1],

Charbel ABDEL NOUR[2], Olivier RIOUL[3], and Catherine DOUILLARD[2]

***Abstract*—DVB-T2 was the first industrial standard deploying rotated and cyclic Q delayed (RCQD)modulation to improve performance over fading channels. This enablesimportantgains compared toconventional quadrature amplitude modulations(QAM) under severe channel conditions.However, the corresponding demodulation complexitystill prevents its use forwider applications. This paper proposes several rotation angles for different QAM constellations anda corresponding low-complexity detection method. Results show that the proposed solution simplifies both the transmitter and the receiver with often betterperformancethan the proposed angles in DVB-T2. Compared with the lowest complexity demappers currently used in DVB-T2, the proposed solution achieves an additional reduction bymore than 60%.**

## I. Introduction

THE DVB-T2standard[1]clearly presents improvedperformance when compared to its predecessor - DVB-T [2]over highly attenuated or erased frequency selective channels. This is partly due to the adoption ofthe rotated and cyclic Q delayed (RCQD) quadrature amplitude modulations (QAM)[3].The idea of the RCQD modulation is to correlate the in-phase (I) and the quadrature (Q) components of the conventional QAM signals by rotation and then to introduce different fading attenuations to I and Q components by distributing these components on different subcarriers.

Due to different attenuations over theI and Q components, decorrelation based methods such as zero forcing (ZF) or minimum mean square error (MMSE) demapper[5]are not optimum anymore. Instead, two dimensional (I and Q)demappers (2D-DEM)are required[6]-[10].For high order constellations such as 64-QAM or 256-QAM, the computational complexity of a 2D-DEM has a non-negligible impact on the receiver design.

The rotation angles of the RCQD modulations of DVB-T2 jointly considered the codes and the modulations, and were selected by applying different criteria as described in[3],[4]. However, these angles were designed to achieve a compromise between performance over fading channels with and without erasures. They do not aim at consideringthe receiver complexity.

Differently from previous works[3],[11]-[17], this paper aims at reducingthe RCQD demapping complexity and proposes a series of rotation angles for different orders of QAM signals with interesting structural properties. Based on these structural properties, a dedicated low-complexity max-log based "sphere-demapper" is designed.Radically different from the sphere-decoders used for MIMO detection[18], the radius of the proposed "sphere-demapper" implies the exact amount of involved constellation points;it also guaranteesthe soft demapping operation to be successfully performed. Moreover, in the worst case, the proposed sphere-demapper achieves almost the same performance as the full-complexity max-log demapper with operational complexity being much less than the demappers proposed in the literature.

The remainder of the paper is organized as follows.The system model and the soft demapping process are introduced in sectionII. The proposed modulation and demodulation solution is detailed in sectionIII. Simulation results, computational complexity and performance comparisons are given in sectionIV. SectionV concludes the proposal.

## II. System Model

### A. Rotated and Cyclic Q Delayed Constellations

A conventional square $M$ -QAM symbol $s$ (where $\sqrt{M}$ is an integer) can take values from the following $\mathbf{S}_c$ set:

$$\mathbf{S}_c = \left\{ s = s_I + j s_Q \,\middle|\, s_I, s_Q \in \mathbf{A}_c \right\}, \tag{1}$$

where $\mathbf{A}_c$ is defined as:

$$\mathbf{A}_c = \left\{ \beta_s \left( -\sqrt{M} + 1 + 2p_m \right) \,\middle|\, p_m \in \left\{ 0, 1, \cdots, \sqrt{M} - 1 \right\} \right\}; \tag{2}$$

$m$ can either be I (in-phase component) or Q (quadrature component) and $\beta_s$ is a normalization factor[1](*e.g.*, for 64-QAM $\beta_s = 1/\sqrt{42}$ and for 256-QAM $\beta_s = 1/\sqrt{170}$ ).

In order to obtain a RCQD constellation, the conventional

Manuscript received September29, 2014. This work was supported in part by the ANRGreencocom project.

[1] JianxiaoYANG, KaiWAN and BenoitGELLER are with Department U2IS of ENSTA-ParisTech, 828 Boulevard des Maréchaux, 91120, Palaiseau, France (e-mail: jianxiao.yang, kai.wan, benoit.geller@ensta-paristech.fr);

[2] Charbel ABDEL NOURand Catherine DOUILLARD are with Lab-STICC UMR CNRS 3192, and they are also with ELEC - Dépt. Electronique, Institut Mines Télécom-Télécom Bretagne, CS 83818, 29238 BREST cedex 3, France (e-mail: charbel.abdelnour, catherine.douillard @telecom-bretagne.eu).

[3] Olivier RIOUL is with COMELEC LTCI of Telecom ParisTech, 46 rue Barrault, 75634 Paris Cedex 13, France(e-mail: olivier.rioul@enst.fr).

square symbol is first rotated by an angle $\theta$ to obtain a rotated symbol $z = s \cdot \exp(j\theta) = z_I + jz_Q$ ,where $z_I$ and $z_Q$ are:

$$\begin{cases} z_I = s_I \cos\theta - s_Q \sin\theta \\ z_Q = s_I \sin\theta + s_Q \cos\theta. \end{cases} \tag{3}$$

Then the real part of symbol $z$ and the imaginary part of another previously delayed symbol $z'$ build the symbol to be transmitted $x$ on a given subcarrier, *i.e.*, $x = \mathrm{Re}(z) + j\,\mathrm{Im}(z')$ .

At the receiver side, the components $\mathrm{Re}(z)$ and $\mathrm{Im}(z)$ of symbol $z$ areattenuated differently.Let $h_I \geq 0$ and $h_Q \geq 0$ denote theRayleigh distributed fading coefficients ofthe two subcarriers where symbol $z$ has been transmitted.The observed symbol $y = y_I + jy_Q$ received by the demappercan beexpressed as:

$$y_I + jy_Q = (h_I z_I + n_I) + j(h_Q z_Q + n_Q), \tag{4}$$

where $n = n_I + jn_Q$ represents a zero-mean circularly symmetric complex Gaussian noise term with variance $\sigma_n^2$ .

## B. Soft-DemappingAlgorithm

The maximum likelihood (ML) soft demapper[21]requires the computation of the log likelihood ratio (LLR)corresponding to each mapped bit $b_i$ (with $i = 0,1,\cdots,\log_2 M - 1$)such that:

$$\begin{aligned} \mathrm{LLR}(b_i) &= \ln\left( \sum_{z \in \mathbf{Z}_c(b_i=0)} \exp\left( -\frac{1}{\sigma_n^2}\left( (y_I - h_I z_I)^2 + (y_Q - h_Q z_Q)^2 \right) \right) \right) \\ &\quad - \ln\left( \sum_{z \in \mathbf{Z}_c(b_i=1)} \exp\left( -\frac{1}{\sigma_n^2}\left( (y_I - h_I z_I)^2 + (y_Q - h_Q z_Q)^2 \right) \right) \right) \\ &= \ln\left( \sum_{z \in \mathbf{Z}_c(b_i=0)} \exp\left( -\frac{d(y,z \mid h_I,h_Q)}{\sigma_n^2} \right) \right) \\ &\quad - \ln\left( \sum_{z \in \mathbf{Z}_c(b_i=1)} \exp\left( -\frac{d(y,z \mid h_I,h_Q)}{\sigma_n^2} \right) \right) \end{aligned} \tag{5}$$

where $\mathbf{Z}_c(b_i = b)$ denotes the subset of $\mathbf{Z}_c$ that contains all constellation points associated with $b_i = b$ and $b = \{0,1\}$ , and $d(y,z \mid h_I,h_Q)$ is the square of the 2D Euclidean distance between the observationsymbol $y$ and the considered constellation symbol $z$ of $\mathbf{Z}_c$ in (3)with known channel attenuations $h_I$ and $h_Q$ :

$$d(y,z \mid h_I,h_Q) = (y_I - h_I z_I)^2 + (y_Q - h_Q z_Q)^2 . \tag{6}$$

The LLR computation of (5)requires the exploration of a signal space containing all the possible $M$ complex-valued constellation points.Asuboptimal soft-demappingsolution with a negligible loss can be obtained by applyingthe max-log approximation [6]over (5):

$$\begin{aligned} \mathrm{LLR}(b_i) &\approx \max_{z \in \mathbf{Z}_c(b_i=0)} \left\{ -\frac{d(y,z \mid h_I,h_Q)}{\sigma_n^2} \right\} - \max_{z \in \mathbf{Z}_c(b_i=1)} \left\{ -\frac{d(y,z \mid h_I,h_Q)}{\sigma_n^2} \right\} \\ &= \min_{z \in \mathbf{Z}_c(b_i=1)} \left\{ \frac{1}{\sigma_n^2} d(y,z \mid h_I,h_Q) \right\} - \min_{z \in \mathbf{Z}_c(b_i=0)} \left\{ \frac{1}{\sigma_n^2} d(y,z \mid h_I,h_Q) \right\} \end{aligned} \tag{7}$$

Note that performing LLR computations by(7)is equivalent to finding the optimum point with global minimum distance among all candidates and then finding $\log_2 M$ minimum distances with one bit information complementary to the global optimum point.

## C. Optimum 2D Joint Detection

Designating $h_{1,1} = h_I \cos\theta$ , $h_{1,2} = -h_I \sin\theta$ , $h_{2,1} = h_Q \sin\theta$ , and $h_{2,2} = h_Q \cos\theta$ , the distance metric $d(y,z \mid h_I,h_Q)$ in (6) can be further expanded as follows:

$$\begin{aligned} d(y,z \mid h_I,h_Q) &= (y_I - h_I \cos\theta s_I + h_I \sin\theta s_Q)^2 \\ &\quad + (y_Q - h_Q \sin\theta s_I - h_Q \cos\theta s_Q)^2 \\ &= (y_I - h_{1,1}s_I - h_{1,2}s_Q)^2 + (y_Q - h_{2,1}s_I - h_{2,2}s_Q)^2 \\ &= (y_I)^2 + (y_Q)^2 - \frac{(y_I h_{1,1} + y_Q h_{2,1})^2}{(h_{1,1}^2 + h_{2,1}^2)} - \frac{(y_I h_{1,2} + y_Q h_{2,2})^2}{(h_{1,2}^2 + h_{2,2}^2)} \\ &\quad + (h_{1,1}^2 + h_{2,1}^2)\left[ s_I - \frac{(y_I h_{1,1} + y_Q h_{2,1})}{(h_{1,1}^2 + h_{2,1}^2)} \right]^2 \\ &\quad + (h_{1,2}^2 + h_{2,2}^2)\left[ s_Q - \frac{(y_I h_{1,2} + y_Q h_{2,2})}{(h_{1,2}^2 + h_{2,2}^2)} \right]^2 \\ &\quad - 2(h_{1,1}h_{1,2} + h_{2,1}h_{2,2}) s_I s_Q . \end{aligned} \tag{8}$$

Note that the first four terms are independent from $s_I$ and $s_Q$ , the fifth and the sixth terms are distances dependent respectively on $s_I$ and $s_Q$ , however, the last term is a cross-correlated term related to both $s_I$ and $s_Q$ . (8) indicatesthat the minimum distance $\min\{d(y,z \mid h_I,h_Q)\}$ cannot be obtained with independent decisions over $s_I$ and $s_Q$ due to the last cross-correlated term at the right-hand side of(8). Therefore computing the maximum likelihood (ML) or the max-log solution requires a 2D joint detection. This implies a $O\{M^2\}$ level complexity for the demapperwhich can be reduced by the adoption of a carefully designedrotation angle as proposed next.

# III. New Signal Space Diversity Solution

## A. Properties of Rotated Constellation with RotationAngle $\theta = \arctan(1/\sqrt{M})$

We propose the rotation angle $\theta = \arctan(1/U)$ for the RCQD, where it should be mentioned that $U = \sqrt{M}$ is an integer value. From (1)-(3), thecomponents $z_I$ and $z_Q$ of the rotated constellation can then be expressed as:

$$\begin{cases} z_I = \left( \frac{-M + 1}{2} + \left( \sqrt{M}\, p_I + \left( \sqrt{M} - 1 - p_Q \right) \right) \right) 2\beta_s \sin\theta \\ z_Q = \left( \frac{-M + 1}{2} + \left( \sqrt{M}\, p_Q + p_I \right) \right) 2\beta_s \sin\theta, \end{cases} \tag{9}$$

where $p_I$ and $p_Q$ are integers given in (2).

This rotation angle introduces the following interesting properties:

**Property 1:** The values $z_I$ and $z_Q$ projected by the rotated constellation $z$ are uniformly distributed along the I-axis and the Q-axis.The minimum distance between any projected consecutive points is $2\beta_s \sin\theta$ , *i.e.*,

$$d_{1D,\min} = \min_{z_I \neq z_I'} |z_I - z_I'| = \min_{z_Q \neq z_Q'} |z_Q - z_Q'| = 2\beta_s \sin\theta \ .$$

**Property 2:** Since $p_I$ and $p_Q$ take integer values from $0$ to the integer $(\sqrt{M}-1)$ (see (1) and (2)), the terms $(\sqrt{M}\,p_I + (\sqrt{M}-1-p_Q))$ and $(\sqrt{M}\,p_Q + p_I)$ in (9)can take $M$ values.Therefore,these latter canbe treated as two-digit numbers, each digit being able to take $\sqrt{M}$ values; consequently from (3), $z_I$ and $z_Q$ can identically take $M$ values. Moreover, without considering the minimum distance factor $2\beta_s \sin\theta$, the most and least significant digits of $z_I$ are respectively $p_I$ and $p_Q$ , while the most and least significant digits of $z_Q$ are respectively $p_Q$ and $p_I$ .

Reciprocally, one can determine $(p_I, p_Q)$ for any given value $z_I$ or $z_Q$ with the following property.

**Property 3:**From (9), one given value $z_I$ uniquely determines one pair $(p_I, p_Q)$:

$$\begin{cases} p_I = \left\lfloor \dfrac{T_I}{\sqrt{M}} \right\rfloor \\ p_Q = \sqrt{M} - 1 - (T_I - \sqrt{M}\,p_I), \end{cases} \tag{10}$$

as well asone given value $z_Q$ uniquely determines one pair $(p_I, p_Q)$:

$$\begin{cases} p_Q = \left\lfloor \dfrac{T_Q}{\sqrt{M}} \right\rfloor \\ p_I = (T_Q - \sqrt{M}\,p_Q), \end{cases} \tag{11}$$

where $T_I$ and $T_Q$ are shown as below:

$$T_I = \frac{z_I}{d_{1D,\min}} + \frac{1}{2}(M-1) \tag{12}$$

$$T_Q = \frac{z_Q}{d_{1D,\min}} + \frac{1}{2}(M-1). \tag{13}$$

Note that the values $T_I$ and $T_Q$ are integers belonging to $[0, M-1]$. Moreover, the integer division by $\sqrt{M}$ (*resp*. the multiplication by $\sqrt{M}$ ) is not implemented by a divider (*resp*. a multiplier), instead, they can be implemented by $\log_2 \sqrt{M}$ right-shifts(or left-shifts).

**Property 4:** Any $\sqrt{M}$ consecutive values of $T_I$ (*resp*. $T_Q$ ) contains exactly all possible values of $p_Q$ (*resp*. $p_I$ ) and each value occurs once.

This property is due to fact that $p_Q$ is the least significant digit of $T_I$ while $p_I$ is the least significant digit of $T_Q$ ;Every consecutive $\sqrt{M}$ $T_I$ or $T_Q$ values contain every integer values from 0 to $\sqrt{M}-1$ .

To synthesize the previous properties, one can find the 2D point $z$ from either its realpart or its imaginary part: given $z_I$ (*resp*. $z_Q$ ) from (12) (*resp*.(13)) (see Fig.1 for QPSK constellation), one first finds $T_I$ (*resp*. $T_Q$ ) (seeFig.2), then finds $(p_I, p_Q)$ from (10) (*resp*.(11)) (seeFig.3), and return $z$ from (9) (see Fig.1).

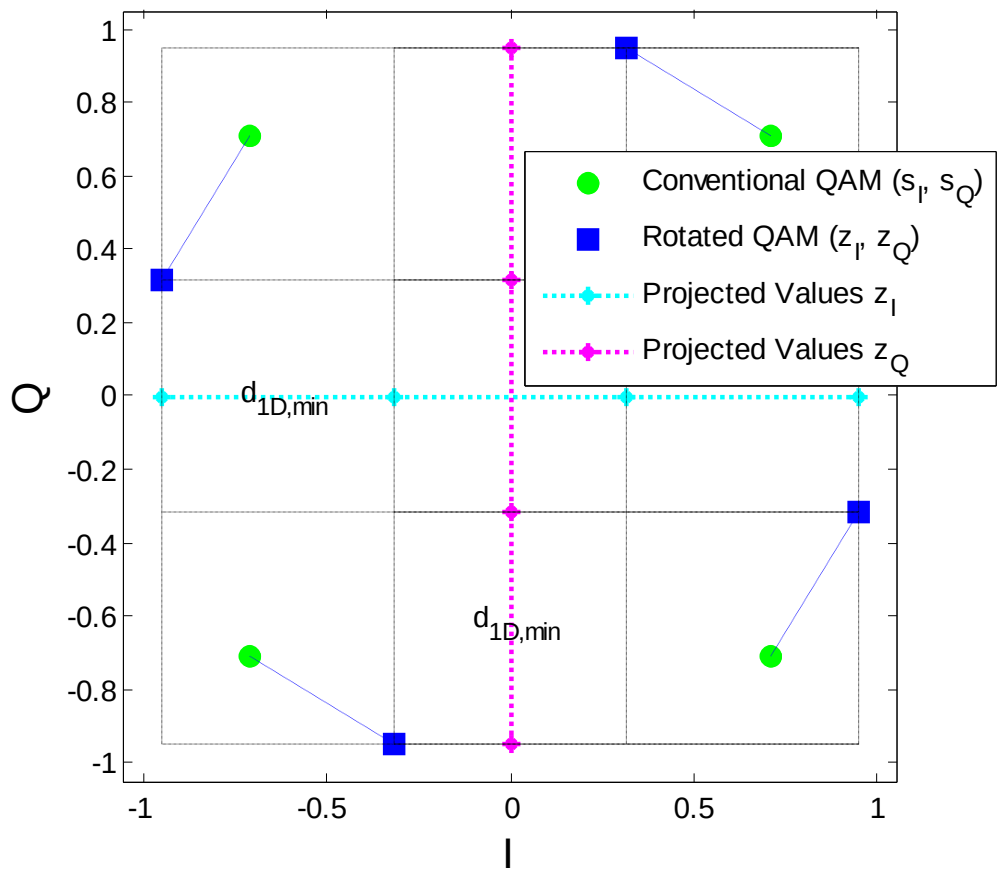

Fig.1. Conventional QPSK $(s_I, s_Q)$ , rotated QPSK $(z_I, z_Q)$ , projected values of rotated QPSK signal and $d_{1D,\min} = 2\beta_s \sin\theta$ .

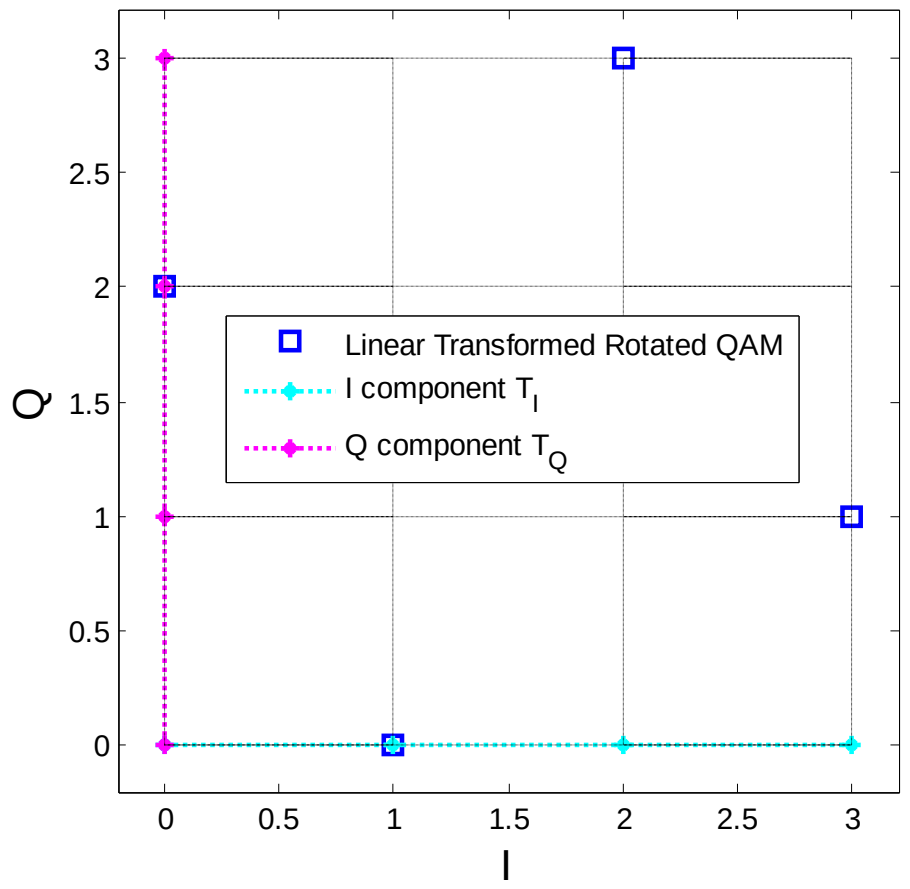

Fig.2. Linear transformed rotated QPSK $(T_I, T_Q)$ from $(z_I, z_Q)$ by using (12)and (13).

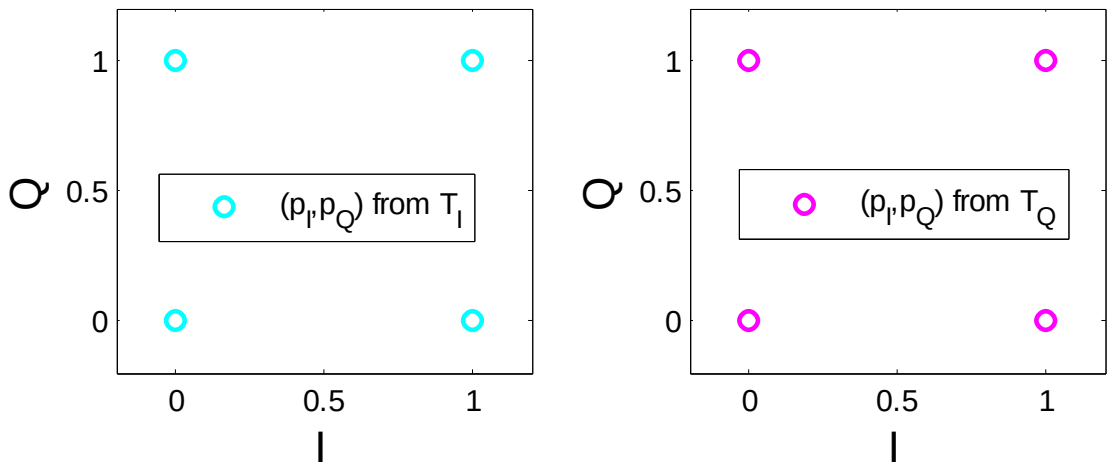

Fig.3. $(p_I, p_Q)$ obtained from either $T_I$ or $T_Q$ by using (10)or (11).

At the receiver side, the observations $y_I$ and $y_Q$ can be equalized as:

$$\begin{aligned} Y_m &= \frac{y_m}{d_{1D,\min} h_m} + \frac{1}{2}(M-1) \\ &= \frac{h_m z_m}{d_{1D,\min} h_m} + \frac{1}{2}(M-1) + \frac{n_m}{d_{1D,\min} h_m} \\ &= T_m + \frac{n_m}{d_{1D,\min} h_m}, \end{aligned} \tag{14}$$

where $m$ can either be I or Q, and $Y_m$ can either be $Y_I$ or $Y_Q$ . From (12)and (14), one has:

$$(y_I - h_I z_I)^2 = d_{1D,\min}^2 \left(h_I (Y_I - T_I)\right)^2 \tag{15}$$

and similarly from (13) and (14):

$$(y_Q - h_Q z_Q)^2 = d_{1D,\min}^2 \left(h_Q (Y_Q - T_Q)\right)^2 . \tag{16}$$

In this way, the distance metric $d\left(y,z \mid h_I,h_Q\right)$ of (6) can be modified as:

$$d\left(y,z \mid h_I,h_Q\right) = d_{1D,\min}^2\left(\left(h_I\left(Y_I - T_I\right)\right)^2 + \left(h_Q\left(Y_Q - T_Q\right)\right)^2\right). \quad (17)$$

Note that (17) simplifies the evaluation of the distance metric (and the corresponding soft demappingin (7)), as the only varying parameters $T_I$ and $T_Q$ are now integer valued.

## B. Low-complexity Sphere-Demapper

Let $\tilde{T}_I$ (*resp.* $\tilde{T}_Q$) be the estimated value of $T_I$ and (*resp.* $T_Q$), obtained by rounding the equalized value $Y_I$ (*resp.* $Y_Q$) of (14)to an integer. From (10) (*resp.* (11)), one can find the only local closest point to $Y_I$ (*resp.* $Y_Q$), and ML point $z$ in the rotated constellation can only be found when $\tilde{T}_I$ and $\tilde{T}_Q$ are pointing to the same pair $\left(p_I,p_Q\right)$, *i.e.*, $\left(\tilde{p}_I\left(\tilde{T}_I\right),\tilde{p}_Q\left(\tilde{T}_I\right)\right) = \left(\tilde{p}_I\left(\tilde{T}_Q\right),\tilde{p}_Q\left(\tilde{T}_Q\right)\right)$ .When $\tilde{T}_I$ and $\tilde{T}_Q$ do not lead to the same constellation point, in order to locate the global optimum point with minimum distance to $\left(Y_I,Y_Q\right)$ (*i.e.*, minimizing (17)), one can increase the observation candidate number by searching regions centered around $Y_I$ and $Y_Q$ with radius $d$ . This region can be expressed as:

$$\tilde{\mathbf{T}}\left(Y_m\right) = \begin{cases} \left[0, 2d-1\right], & \text{if } Y_m < d, \\ \left[M-2d, M-1\right], & \text{if } Y_m \geq M-d, \\ \left[\lfloor Y_m \rfloor - d + 1, \lfloor Y_m \rfloor + d\right] & \text{else,} \end{cases} \quad (18)$$

where $m$ can either be I or Q, and $Y_m$ can either be $Y_I$ or $Y_Q$ .

Note that points within $\tilde{\mathbf{T}}\left(Y_m\right)$ are uniformly distributed along the axis $m$ with unit interval, *i.e.*, 1.With the given radius $d$ , $\tilde{\mathbf{T}}\left(Y_m\right)$ in (18) computes exactly $2d$ points and thus searchingthe global optimum point requires a search to be performed among $4d$ points within $\tilde{\mathbf{T}}\left(Y_I\right)$ and $\tilde{\mathbf{T}}\left(Y_Q\right)$ .

Differently from the classical sphere-decoder proposed in [18], $Y_m$ in (14) are equalized observations instead of the receivedunequalized observations $y_I$ and $y_Q$ in [18]. Moreover, the integer radius $d$ means that $2d$ points are to be found within each region $\tilde{\mathbf{T}}\left(Y_m\right)$ , whereas the radius in [18]gives no indication on the number of points involved in the sphere decoding.

Due to the useful property 4, the radius used by the sphere demappingalgorithm is $\sqrt{M}/2$ . Therefore, the sphere demappingalgorithm can be summarized into four steps:

1)Transform linearly the received signals $y_I$ and $y_Q$ into $Y_I$ or $Y_Q$ by using (14);

2) Compute the regions $\tilde{\mathbf{T}}\left(Y_I\right)$ and $\tilde{\mathbf{T}}\left(Y_Q\right)$ centered around $Y_I$ or $Y_Q$ by using (18);

3) Compute the Euclidean distance metrics within the $4d$ points of step 2) by using (6);

4)Perform LLR computations by(7), *i.e.*,finding the global minimum distance among all candidates andthen finding $\log_2 M$ minimum distances with one bit information complementary to the global optimum point.

The detailed analysis on the complexity of this algorithm is evaluated in terms of number of candidate points (CPs), real multiplications (RMs), real comparisons (RCs), real inversions (RIs) and real sums (RSs), where RS can either be real additions or real subtractions.

1) In step 1, the terms $1/d_{1D,\min}^2$ and $(M-1)/2$ are constants and do not need repeated computations, therefore the transformations of (14)require 2RSs, 2RMs and 2RIs.

2) In step 2, according to (18), the selection of two regions $\tilde{\mathbf{T}}\left(Y_I\right)$ and $\tilde{\mathbf{T}}\left(Y_Q\right)$ needs 4 RCs. Generating all points within the two regions requires another $2\sqrt{M}$ RSs.

3) In step 3, computing one 2D Euclidean distance metric $\left(h_I\left(Y_I - T_I\right)\right)^2 + \left(h_Q\left(Y_Q - T_Q\right)\right)^2$ requires 4 RMs and 3RSs. Since there are $2\sqrt{M}$ distances to compute, $8\sqrt{M}$ RMs and $6\sqrt{M}$ RSs are required in total.

4) In order to perform LLR computations asin(7), one needs to find the global minimum distance among all $2\sqrt{M}$ candidates and this requires $2\sqrt{M}-1$ RCs. Moreover, one also has to compute $\log_2 M$ minimum distances with one bit information complementary to the global optimum point. Since each complementary bit needs $2\sqrt{M}-2$ RCs, finding complementary points needs $\left(2\sqrt{M}-2\right)\log_2 M$ in total.After finding the minimum distances, one also needs additional $\log_2 M$ RMs and $\log_2 M$ RSs.

Therefore, performing LLR computations in (7) requires $2\sqrt{M}$ CPs, $8\sqrt{M}+2+\log_2 M$ RMs, 2 RIs, $8\sqrt{M}+2+\log_2 M$ RSs and $5+\left(2\sqrt{M}-2\right)\left(1+\log_2 M\right)$ RCs.

# IV. Simulation and Results

This section is divided into two subsections.The first subsectioncompares the bit error rate (BER) of the new RCQD solution with rotation angle $\theta = \arctan\left(1/\sqrt{M}\right)$ with the BER of the classical RCQD used in DVB-T2for 16-QAM and 64-QAM over fading and fading erasure channels. The second subsection compares the proposed sphere demapping algorithm with other currently proposed algorithms for 256-QAM signal in terms of performance and complexity. It should be mentioned that the 256-QAM is the only constellation using rotation angle $\theta = \arctan\left(1/\sqrt{M}\right)$ in DVB-T2.

In these simulations, perfect synchronization and channel estimation are assumed, which is different from the practical case [19],[20].BERs are obtained with the DVB-T2 LDPC code with length 64800 and rate 4/5 with min-sum decoding applying 25 iterations. Performanceisevaluated over Rayleigh fading channels with and without erasure events as defined in[21].

## A. The New RCQD Solution with rotation angle $\theta = \arctan\left(1/\sqrt{M}\right)$

Fig.4 and Fig.5 compare the BER performance of the proposed rotation angles with the DVB-T2 rotation angles over the Rayleigh fading channel. It can be observed that the performance of the RCQD with the proposed angle loses 0.05 dB (*resp.* 0.025 dB) compared with the performance of the RCQD with the DVB-T2 angle for 16-QAM (*resp*. 64-QAM) signal at a BER = $10^{-6}$. The negligible performance loss isdue to the fact that the proposed angles are not jointly designed with the LDPC codesfor a maximization of the coding gain.

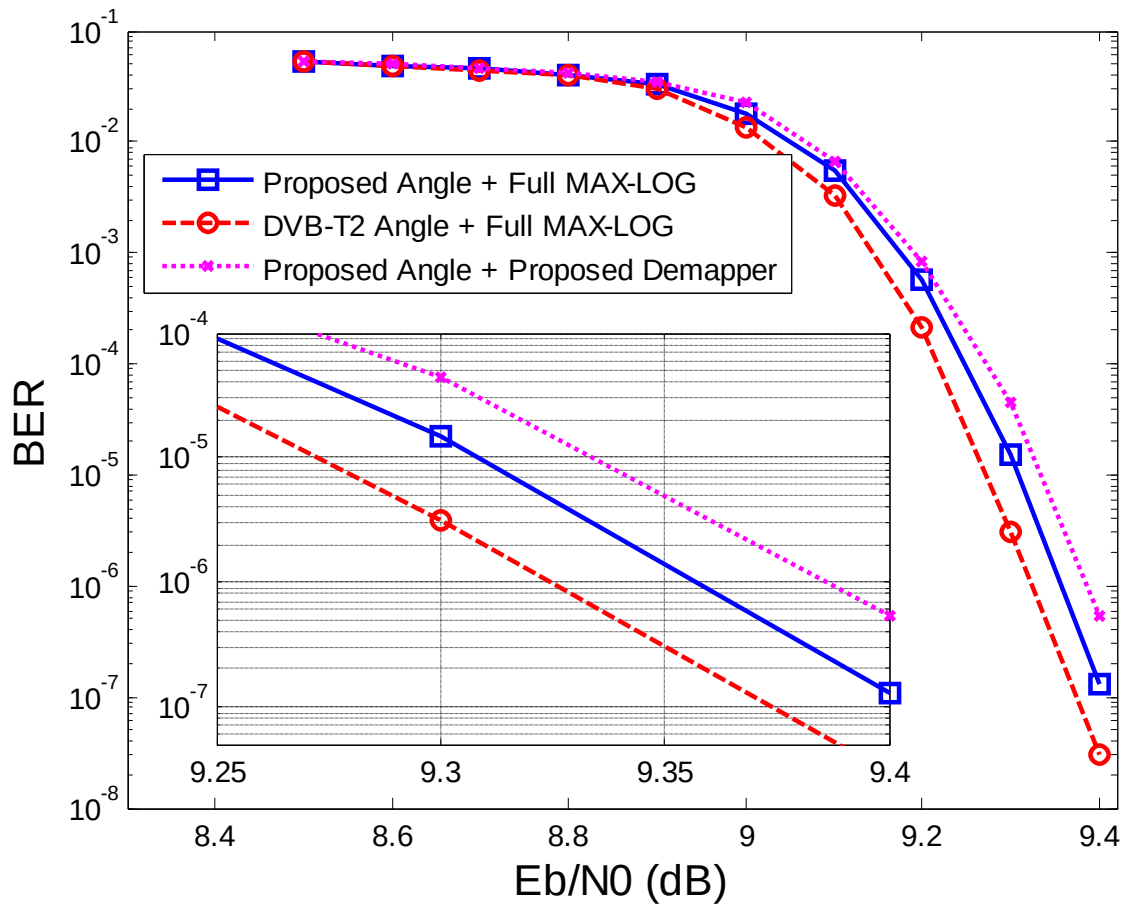

Fig.4. BER comparison of the 16-QAM RCQD solutions with proposed rotation angle (14.0 degree), rotation angle proposed by DVB-T2 (16.8 degree) over a fading channel.

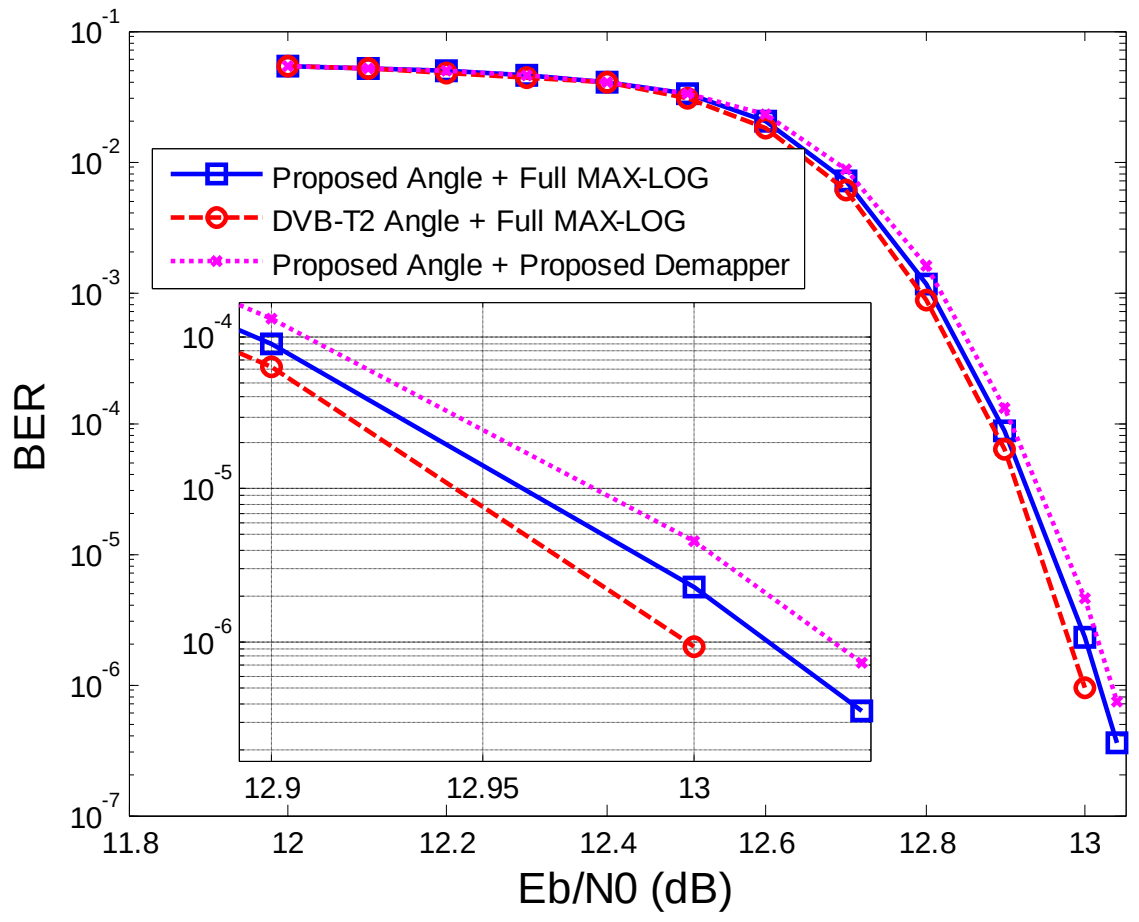

Fig.5. BER comparison of the 64-QAM RCQD solutions with proposed rotation angle (7.1 degree), rotation angle proposed by DVB-T2(8.6 degree) over a fading channel.

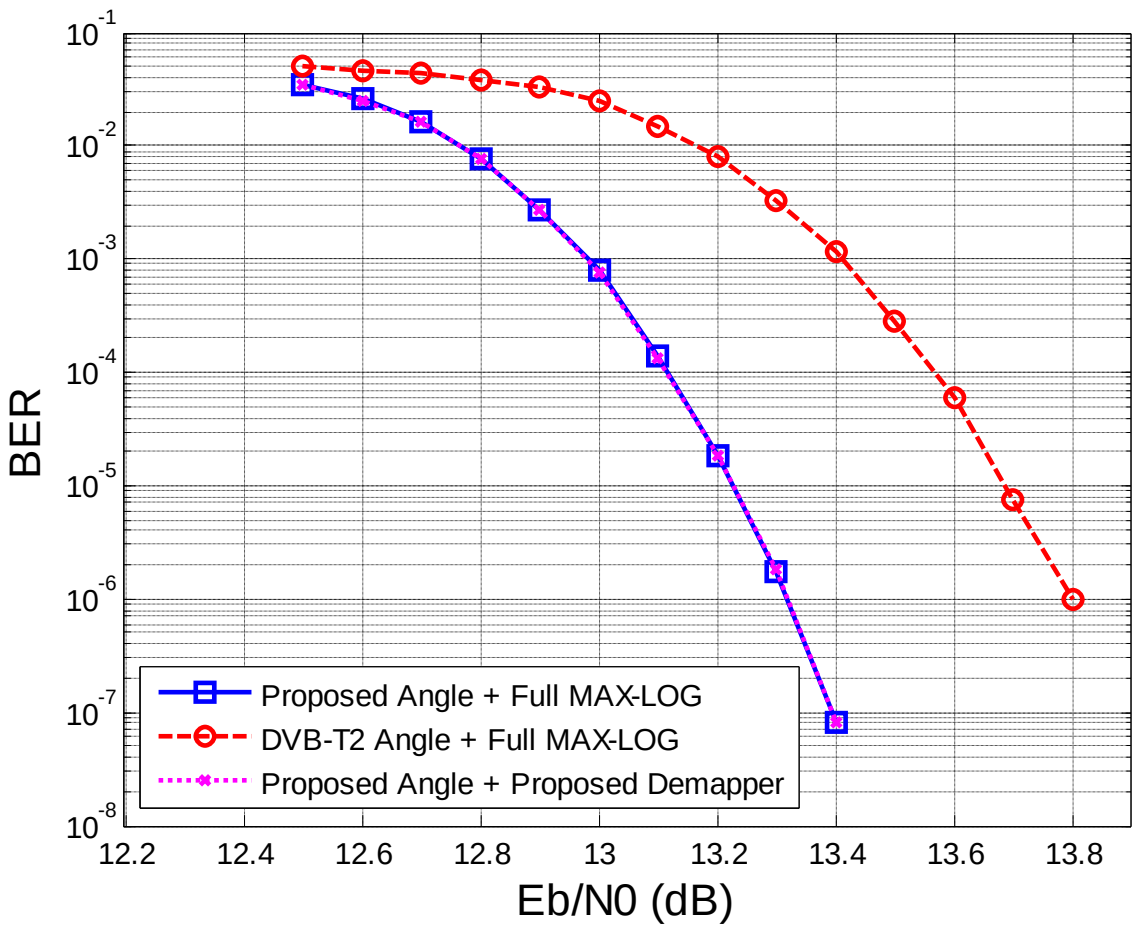

Fig.6. BER comparison of the 16-QAM RCQD solutions with proposed rotation angle, rotation angle proposed by DVB-T2 over a fading channel with 15% of erasure events.

Fig.6 and Fig.7 compare the BER performance over fading erasure channel with a 15% erasure rate. It can be easily noticed that the proposed rotation angle outperforms the original DVB-T2 rotation angle by 0.5 dB (*resp*. 0.75 dB) for 16-QAM (*resp*. 64-QAM) signal at a BER = $10^{-6}$. The larger gain of the erasure channel is because the proposed rotation angle maximizes the minimum interval between two consecutive points projected over real and imaginary axis (see property 1).

Furthermore, the proposed low-complexity sphere based algorithm achieves the same performance as the max-log algorithm with full complexity. The structural properties of the proposed rotation angle not only introduce a simplified demapping algorithm but also leadto improved robustness of the system. Similar results are observed for other erasure rates.

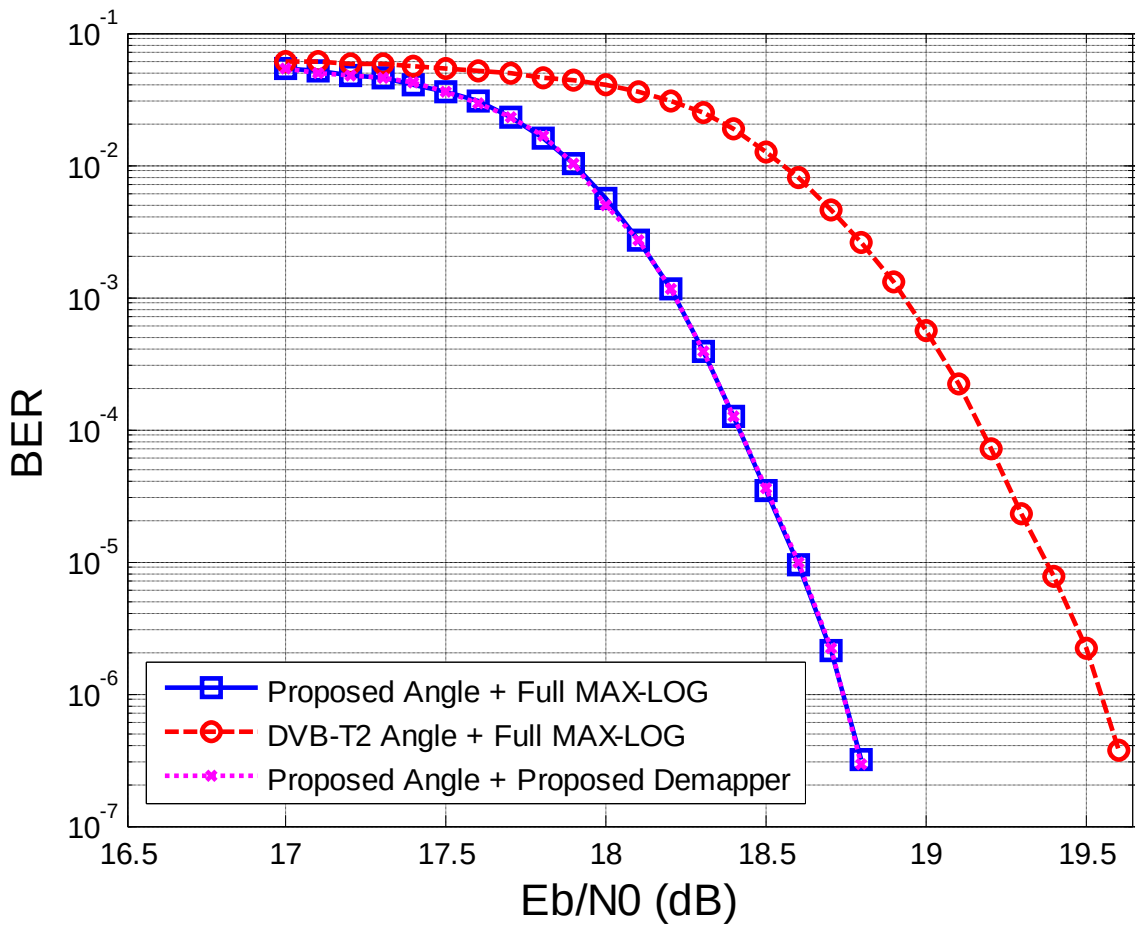

Fig.7. BER comparison of the 64-QAM RCQD solutions with proposed rotation angle, rotation angle proposed by DVB-T2 over a fading channel with 15% of erasure events.

## *B. The Proposed Sphere based Demapping Algorithm*

In this part, the proposed sphere based demapping algorithm is compared with other methods in terms of BER and computation complexity, such as the max-log method (see(5)), the MMSE method [5], the sub-region method [6], and the PD-DEM method[10].

Since the rotation angle of the DVB-T2 RCQD 256-QAM signal satisfies $\theta = \arctan\left(1/\sqrt{M}\right)$, the proposed demapper can also be applied to this case directly. TABLE I gives the complexity comparison in terms of CP, RM, RS, RC and RI(these abbreviations are defined in section III.B) todemapthe RCQD 256-QAM signal.

TABLE I Complexity Comparison of the considered algorithms for the DVB-T2 RCQD 256-QAM constellation.

| Algorithm | CP | RM | RS | RC | RI |
|---|---|---|---|---|---|
| Max-Log | 256 | 1032 | 776 | 2048 | 0 |
| Sub-region | 81 | 332 | 251 | 648 | 0 |
| MMSE | 16 | 64 | 48 | 128 | 6 |
| PD-DEM | 80 | 390 | 279 | 231 | 0 |
| Proposed | 32 | 138 | 138 | 275 | 2 |

Fig.8 and Fig.9 display the BER performance of the various algorithms. Among them, the MMSE method has the lowest computation complexity. However, it achieves the worst performance among all the simulated methods; this is due to the fact that the decorrelation based algorithm is not the optimum method (see the last term in (8)) for the RCQD constellations. Since the other simulated algorithms are 2D joint detection based methods, they are always better than the MMSE algorithm (except the PD-DEM method over a fading channel).

The proposed method achieves almost the same performance as the full complexity max-log algorithm.However, it reduces by 88% the number of CPs, 87% the number of RMs, 82% the number of RSs and 87% the number of RCs with respect to the full complexity max-log algorithm and requires 2 additional RIs. Furthermore, compared with the PD-DEM which is the simplest method among all the considered methods, the proposed demapping algorithm introduces a 60% reduction in number of CPs and 64% reduction in number of RMs.

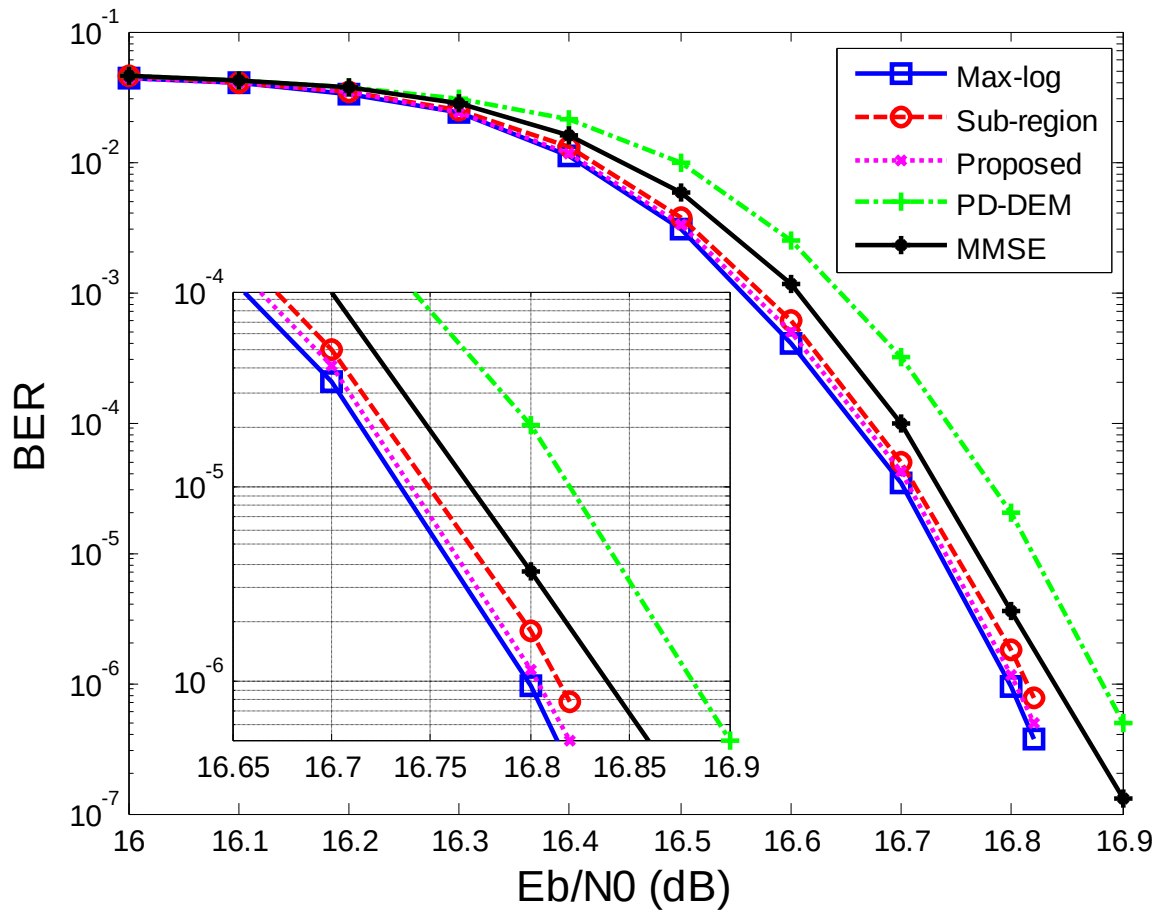

Fig.8. BER comparison of the considered methods for the DVB-T2 RCQD 256-QAM constellation over a fading channel.

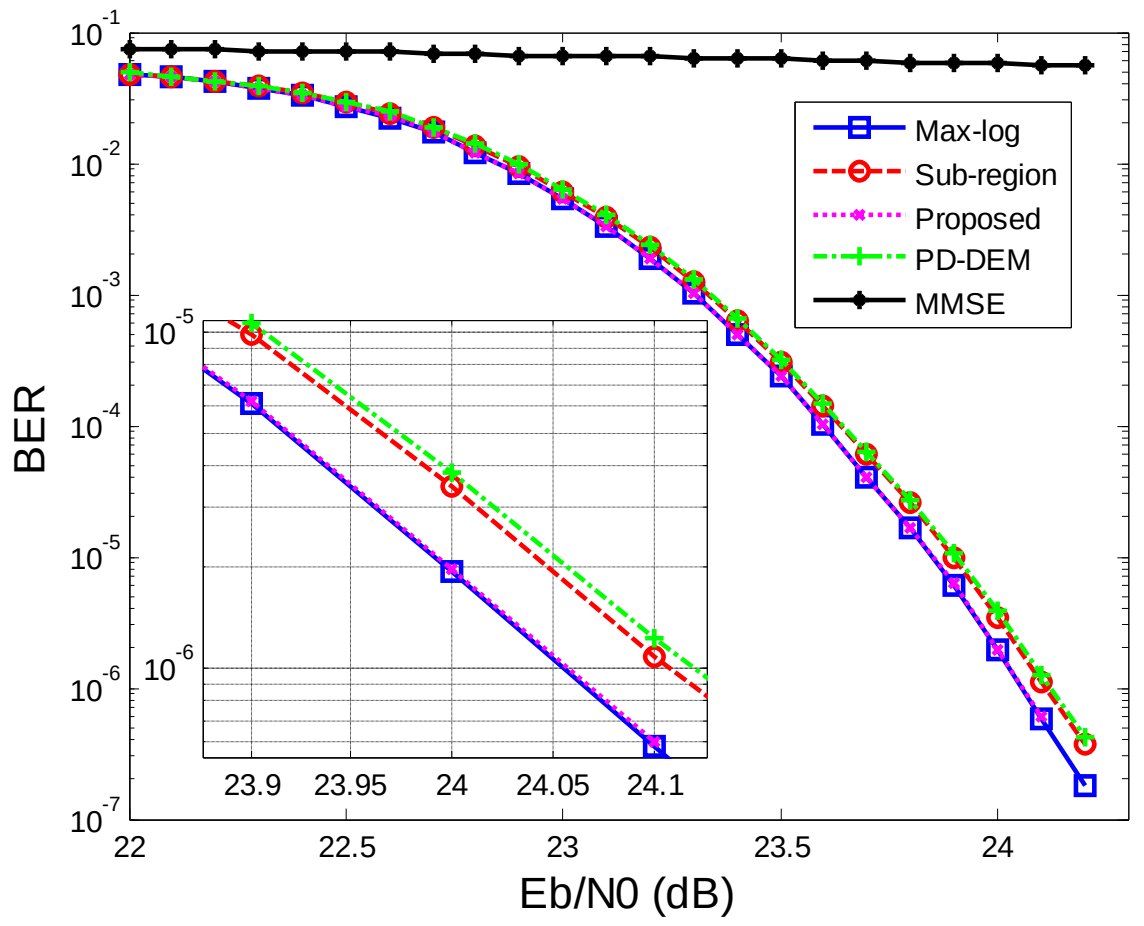

Fig.9. BER comparison of the considered methods for the DVB-T2 RCQD 256-QAM constellation over a fading channel with 15% of erasure events.

## V. Conclusion

This paper proposes a series of rotation angles for different RCQD signals havinginteresting structural properties. Based on these properties, a low-complexity sphere-based max-log demapperis designed. This makes it particularly suited for a hardware implementation. In addition, the proposed solution improves the robustness of the current DVB-T2 system. Due to all these features, this workenables a wider application for the RCQD QAM constellations.

## References

[1] EN 302 755, Digital Video Broadcasting (DVB); Frame structure channel coding and modulation for a second generation digital terrestrial television broadcasting system (DVB-T2), v1.1.1., ETSI, Sept., 2009.

[2] EN300744 V1.6.1, Digital Video Broadcasting (DVB); Framing structure, channel coding and modulation for digital terrestrial television, ETSI, Jan. 2009.

[3] C. Abdel Nour and C. Douillard, "Improving BICM performance of QAM constellation for broadcast applications," *in Proc.Int. Symp. on Turbo Codes and Related Topics*, pp. 55-60, Sept. 2008.

[4] C. Douillard and C.A. Nour, "Rotated QAM Constellations to Improve BICM Performance for DVB-T2," *in Proc. Inter. Symp. on Spread Spectrum Tech. and App., 2008. ISSSTA '08.*,pp.354-359,Aug.2008.

[5] K. Kim, K. Bae, and H. Yang, "One-Dimensional Soft-Demapping using Decorrelation with Interference Cancellation for Rotated QAM Constellations," *IEEE Consum. Commun. and Net. Conf.(CCNC)-Wireless Consum. Commun. and Net.*, pp. 787-791, Jan. 2012.

[6] M. Li, C. Abdel Nour, C. Jégo, and C. Douillard, "Design of rotated QAM mapper/demapper for the DVB-T2 standard," *in Proc.IEEE Workshop on Signal Proc. Systems (SIPS 2009)*, Oct. 2009, pp. 18-23.

[7] M. Li, C.A. Nour, C. Jego, and C. Douillard, "Design and FPGA prototyping of a bit-interleaved coded modulation receiver for the DVB-T2 standard,"*in Proc. of IEEE Workshop on Signal Proc. Systems (SIPS 2010)*, pp.162-167, Oct. 6-8, 2010.

[8] M. Li, C.A. Nour, C. Jego, J. Yang, and C. Douillard, "Efficient iterative receiver for bit-Interleaved Coded Modulation according to the DVB-T2 standard,"*in Proc. of IEEE Intern. Conf. on Acoustics, Speech and Signal Proc. (ICASSP 2011)*, pp. 3168-3171, May 22-27, 2011

[9] D. Pérez-Calderón, V. Baena-Lecuyer, A. C. Oria, P. López, and J. G.Doblado, "Rotated constellation demapper for DVB-T2," *IEEE Electron. Lett.*, vol. 47, no. 1, pp. 31-32, Jan. 2011.

[10] M. Butussi, and S. Tomasin, "Low Complexity Demapping of Rotated and Cyclic Q Delayed Constellations for DVB-T2," *IEEE Wireless Commun. Lett.*, vol.1, pp.81-84, Apr. 2012.

[11] J. Boutros and E. Viterbo, "Signal space diversity: A power- and bandwidth-efficient diversity technique for the Rayleigh fading channel," *IEEE Trans. Inform. Theory*, vol. 44, no. 4, pp. 1453-1467, July 1998.

[12] G. Taricco and E. Viterbo, "Performance of component interleaved signal sets for fading channels," *Electronics Letters*, vol. 32, no. 13, pp. 1170-1172, Apr. 1996.

[13] N. F. Kiyani and J. H. Weber, "Performance analysis of a partially coherent system using constellation rotation and coordinate interleaving," *in Proc. IEEE Globecom*, Dec. 2008, pp. 1-5.

[14] A. Chindapol and J. A. Ritcey, "Bit-interleaved coded modulation with signal space diversity in Rayleigh fading," *in Proc. 33rd Asilomar Conf. Signals, Systems, Computers*, 1999, pp. 1003-1007.

[15] N. F. Kiyani, U. H. Rizvi, J. H. Weber, and G. J. Janssen, "Optimized rotations for LDPC-coded MPSK constellations with signal space diversity," *in Proc. IEEE WCNC*, Mar. 2007, pp. 677-681.

[16] Y. Li, X.-G. Xia, and G. Wang, "Simple iterative methods to exploit the signal-space diversity," *IEEE Trans. Commun.*, vol. 53, no. 1, pp.32-38, Jan. 2005.

[17] M. N. Khormuji, U. H. Rizvi, G. J. Janssen, and S. B. Slimane, "Rotation optimization for MPSK/MQAM signal constellations over Rayleigh fading channels," *in Proc. IEEE ICCS*, Oct. 2006, pp. 1-5.

[18] B. Hassibi and H. Vikalo, "On the Sphere-Decoding Algorithm I. Expected Complexity," *IEEE Trans. Signal Process.*, vol. 53, no. 8, pp. 2806-2818, Aug. 2005.

[19] J. Yang, B. Geller and S. Bay, "Bayesian and Hybrid Cramér-Rao Bounds for the Carrier Recovery under Dynamic Phase Uncertain Channels", *IEEE Transaction on Signal Processing*, pp.667- 680, Feb. 2011.

[20] J. Yang and B. Geller, "Near-optimum Low-Complexity Smoothing Loops for Dynamical Phase Estimation", *IEEE Trans. on Signal Processing*, pp.3704-3711, Sept. 2009.

[21] TR 102 831 V1.1.1, Implementation guidelines for a second generation digital terrestrial television broadcasting system (DVB-T2), ETSI, Oct. 2010.